\documentclass{article}
\usepackage{dcase2024,amsmath,graphicx,url,times,booktabs, tabularx, comment}

\usepackage{hyperref, url}

\usepackage{hhline}
\usepackage{hyperref}
\usepackage{xcolor}
\usepackage{cite}
\usepackage{multirow,url,hyperref}
\usepackage{siunitx}

\usepackage{pifont}
\newcommand{\cmark}{\ding{51}}%
\newcommand{\xmark}{\ding{55}}%
\usepackage{multirow}

\title{Multi-Iteration Multi-Stage Fine-Tuning of Transformers \\ for Sound Event Detection with Heterogeneous Datasets}

\name{Florian Schmid$^1$, Paul Primus$^1$, Tobias Morocutti$^1$,  Jonathan Greif$^{1}$, Gerhard Widmer$^{1,2}$}
\address{$^1$Institute of Computational Perception (CP-JKU), \ $^2$LIT Artificial Intelligence Lab,\\          
        Johannes Kepler University Linz, Austria \\
        \{florian.schmid, paul.primus\}@jku.at\\ 
 }

\begin{document}

\ninept
\maketitle

\begin{sloppy}

\begin{abstract}
A central problem in building effective sound event detection systems is the lack of high-quality, strongly annotated sound event datasets. For this reason, Task 4 of the DCASE 2024 challenge proposes learning from two heterogeneous datasets, including audio clips labeled with varying annotation granularity and with different sets of possible events. We propose a multi-iteration, multi-stage procedure for fine-tuning Audio Spectrogram Transformers on the joint \textit{DESED} and \textit{MAESTRO Real} datasets. The first stage closely matches the baseline system setup and trains a CRNN model while keeping the pre-trained transformer model frozen. In the second stage, both CRNN and transformer are fine-tuned using heavily weighted self-supervised losses. After the second stage, we compute strong pseudo-labels for all audio clips in the training set using an ensemble of fine-tuned transformers. Then, in a second iteration, we repeat the two-stage training process and include a distillation loss based on the pseudo-labels, achieving a new single-model, state-of-the-art performance on the public evaluation set of DESED with a PSDS1 of 0.692. A single model and an ensemble, both based on our proposed training procedure, ranked first in Task 4 of the DCASE Challenge 2024.\footnote{Code: \url{https://github.com/CPJKU/cpjku_dcase24}}.
\end{abstract}



\vspace{-4pt}
\section{Introduction}
\label{sec:intro}
\vspace{-4pt}

The goal of Sound Event Detection (SED) is to identify specific acoustic events and their timing within audio recordings. Reliable SED systems enable applications in numerous domains, for example, in security and surveillance~\cite{radhakrishnan2005audio}, smart homes~\cite{debes2016monitoring}, or health monitoring~\cite{zigel2009method}. A main driver of research in this field is the annual DCASE Challenge, particularly Task 4, which focuses on SED. 

State-of-the-art SED systems are based on deep learning approaches, requiring a substantial amount of annotated data. Their performance is mainly limited by the lack of strongly annotated real-world sound event datasets~\cite{Martinmorato2023maestro}. Hence, previous editions of Task 4 focused on learning from weakly labeled data~\cite{kong2020sound}, semi-supervised learning strategies~\cite{park2021self}, and utilizing synthetic strongly labeled data~\cite{Turpault2019DCASE}. While Task 4 has been based on the DESED dataset~\cite{Turpault2019DCASE} in previous years, the key novelty of the 2024 edition is a unified setup including a second dataset, MAESTRO Real~\cite{Martinmorato2023maestro}. As domain identification is prohibited, the goal is to develop a single system that can handle both datasets despite crucial differences, such as labels with different temporal granularity and potentially missing labels. In fact, because of the lack of strongly annotated, high-quality real-world data, the hope is that learning from two datasets in parallel has a synergetic effect and eventually increases the generalization performance on both datasets. 

The main contributions of this work are as follows: \textbf{(1)} We introduce a multi-iteration, multi-stage training routine for fine-tuning pre-trained transformer models on SED using heterogeneous datasets. \textbf{(2)} We demonstrate that combining fine-tuned transformers -- ATST~\cite{li2022atst}, PaSST~\cite{koutini22passt}, and BEATs~\cite{chen2022beats} -- into a diverse ensemble to generate pseudo-labels, and using these pseudo-labels in a subsequent training iteration, significantly enhances single-model performance, yielding a relative increase of 25.9\% in terms of polyphonic sound detection score~\cite{ebbers2022threshold, bilen2020framework_psds1} (PSDS1) on DESED and 2.7\% in terms of segment-based mean partial area under the ROC curve (mpAUC) on MAESTRO, compared to the baseline system. \textbf{(3)} We conduct an ablation study to analyze the impact of the heterogeneous datasets and design choices related to them.

On DESED, we set a new state of the art on the public evaluation set, raising single-model performance from 0.686~\cite{ebbers2022threshold} to 0.692 in terms of PSDS1. A single model and an ensemble, both based on our proposed training procedure, ranked first in the respective categories in Task 4 of the DCASE Challenge 2024~\cite{cornell2024dcase}.

\vspace{-4pt}
\section{Related Work}
\label{sec:rel}
\vspace{-4pt}


\ \ \ 
\textbf{SED Architectures:} Since the 2018 edition~\cite{Serizel2018dcase}, the baseline system is based on a Convolutional Recurrent Neural Network (CRNN). A large increase in performance happened in the 2023 edition, as the baseline used BEATs~\cite{chen2022beats} embeddings concatenated with CNN embeddings, which were then fed to the RNN, with a relative increase of almost 50\% in PSDS1 score. Top-ranked systems in the 2023 edition improved over the baseline architecture with variations of frequency-dynamic convolution~\cite{nam2022frequency}. Recently, Shao et al.~\cite{shao2024fine} proposed a procedure to fine-tune large pre-trained transformers on the DESED dataset with a two-stage training procedure, establishing a new state of the art. They showed that the key to avoiding overfitting is placing a large weight on the self-supervised losses to take advantage of the larger amount of unlabeled data.

\textbf{Data Augmentation:} As strongly annotated data is very limited, data augmentation is an important strategy to improve the generalization of SED systems. In this regard, Filter-Augment~\cite{nam2022filteraugment} simulates variations in the acoustic environment by applying different weights to frequency bands, forcing the model to extract information from wider frequency regions. Strategies for recording device generalization in Acoustic Scene Classification apply similar frequency weighting mechanisms: Frequency-MixStyle~\cite{Kim22freqm,schmid2022knowledge} mixes the frequency information of two audio clips in the dataset, and Device-Impulse augmentation~\cite{morocutti2023device} convolves an audio clip with an impulse response of a real recording device. Recently, Frequency Warping~\cite{li2022atst}, which stretches or squeezes spectrograms in the frequency dimension, was shown to be an integral part when fine-tuning transformers on the DESED dataset~\cite{shao2024fine}.

\textbf{Pseudo-labels:} Both of the top-ranked approaches in the 2022 and 2023 editions of Task 4 computed pseudo-labels. Ebbers et al.~\cite{ebbers2022pre} use a multi-iteration self-training procedure in which pseudo-labels, predicted by an ensemble, are iteratively refined. Kim et al.~\cite{kim2023semi} employ a two-iterations setup in which strong pseudo-labels for weakly labeled, unlabeled, and AudioSet~\cite{gemmeke17audioset} clips are computed from an ensemble of models from the first training iteration. The computed pseudo-labels are converted into hard labels and used as additional targets in a second training iteration.

\begin{figure}[t!]
\vspace{-15pt}
\centering
{\includegraphics[width=\linewidth]{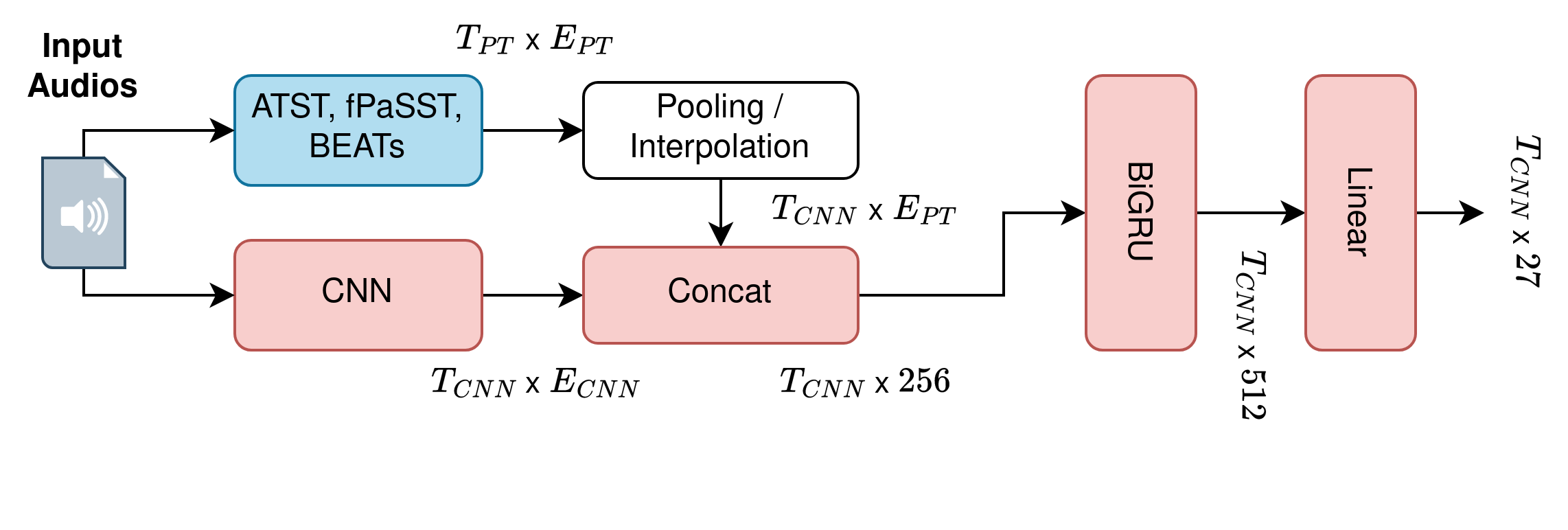}}
\vspace{-35pt}
\caption{Overview of System Architecture. Blue are the pre-trained transformers; the red blocks together comprise the CRNN.}
\label{fig:overview_sys}
\vspace{-15pt}
\end{figure}

\vspace{-4pt}
\section{System Architecture}
\label{sec:arch}
\vspace{-4pt}

Figure~\ref{fig:overview_sys} gives an overview of our SED system. It consists of two independent audio embedding branches (CNN and transformer), the outputs of which are pooled to the same sequence length. A Recurrent Neural Network (RNN) derives strong predictions from these combined sequences. Compared to the baseline~\cite{cornell2024dcase} we experiment with two additional Audio Spectrogram Transformers besides BEATs~\cite{chen2022beats}, namely, ATST~\cite{li2022atst} and PaSST~\cite{koutini22passt}. In addition to adaptive average pooling, we experiment with linear and nearest-exact interpolation to align transformer and CNN sequence lengths. In the following, we briefly describe the transformer models used in our setup. We refer the reader to~\cite{Schmid2024tech_repo} for more details.

\textbf{ATST-Frame}~\cite{li2022atstf}(denoted ATST in the following) was specifically designed to produce a sequence of frame-level audio embeddings. The architecture of ATST is based on the Audio Spectrogram Transformer (AST)~\cite{gongast}; it is pre-trained in a self-supervised manner on AudioSet. In our experiments, we use a checkpoint of ATST that is fine-tuned on the weak labels of AudioSet.

\textbf{fPaSST:} The Patchout faSt Spectrogram Transformer (PaSST)~\cite{koutini22passt} is an improved version of the original AST~\cite{gongast} that shortens the training time and improves the performance via patchout regularization. We adapt PaSST to return frame-level predictions and call the resulting model Frame-PaSST (fPaSST). We pre-train fPaSST on the weakly annotated AudioSet using Knowledge Distillation~\cite{schmid2023efficient}, obtaining a mAP of 0.484.

\textbf{BEATs:} Likewise, BEATs~\cite{chen2022beats} is also based on the AST~\cite{gongast} architecture; it was trained in an iterative, self-supervised procedure on AudioSet, where the BEATs encoder and tokenizer were alternately updated. In our experiments, we rely on the checkpoint of BEATs after the third iteration, where both the tokenizer and the encoder were fine-tuned on the weak labels of AudioSet.

\vspace{-4pt}
\section{Training Pipeline}
\label{sec:training_setup}
\vspace{-4pt}

In this section, we describe the pre-training routine on AudioSet strong and how the pre-trained models are fine-tuned on the Task~4 datasets in the proposed multi-iteration, multi-stage training procedure. An overview of the full training pipeline is shown in Figure~\ref{fig:overview_train}. The full system architecture, depicted in Figure~\ref{fig:overview_sys}, is involved in all iterations and stages of Figure~\ref{fig:overview_train}. The pre-trained transformers (blue block in Figure~\ref{fig:overview_sys}) are used as frozen audio embedding models in Stage 1 and fine-tuned together with the CRNN (red blocks in Figure~\ref{fig:overview_sys}) in Stage 2. The pseudo-labels are generated from an ensemble after Iteration 1 and used as additional prediction targets in Stage 1 of Iteration 2.  In the following, we abbreviate Iteration \{1,2\} and Stage \{1,2\} as \textit{I\{1,2\}} and \textit{S\{1,2\}}, respectively.    

\begin{figure}[t!]
\vspace{-15pt}
\centering
{\includegraphics[width=1.0\linewidth]{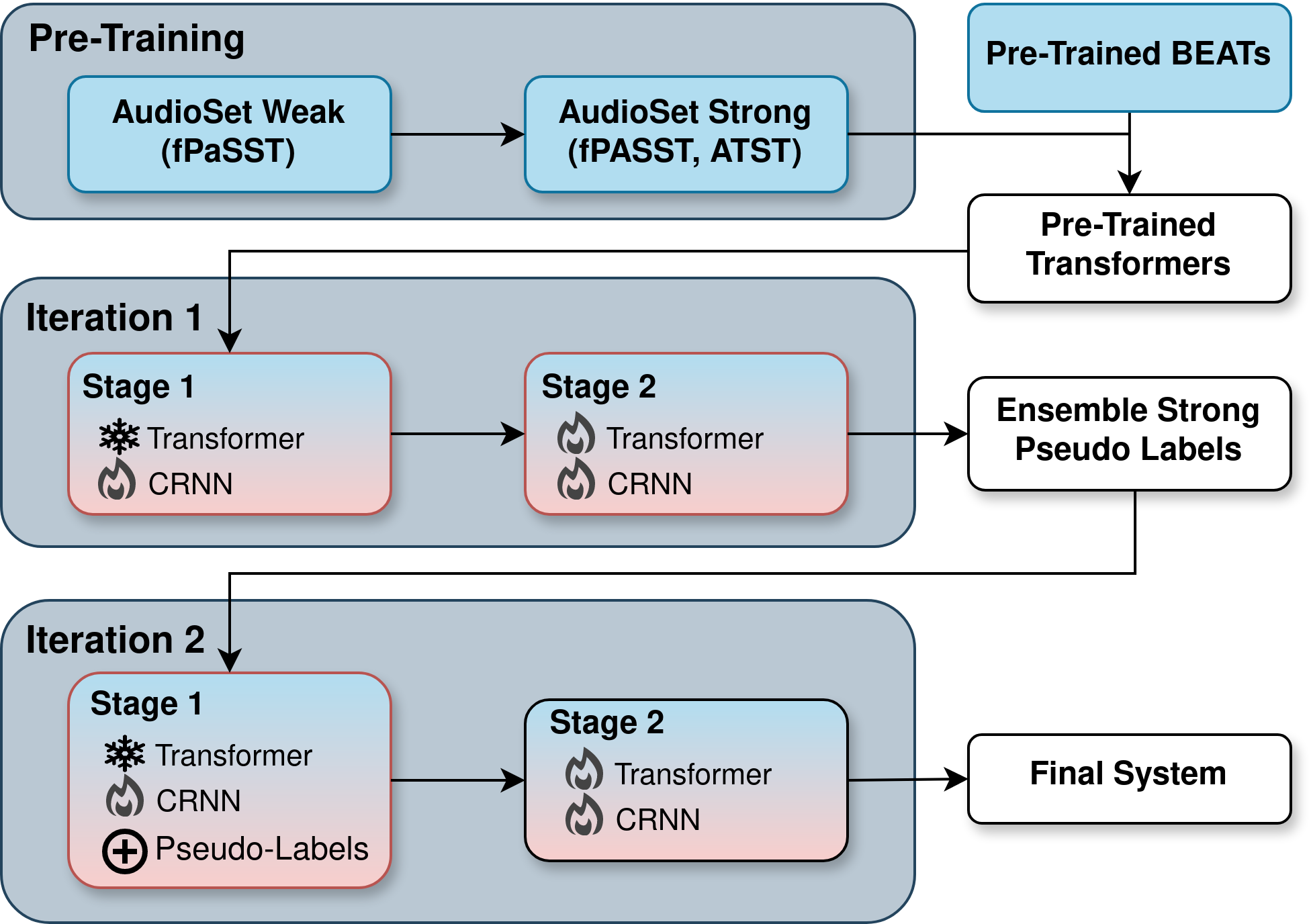}}
\caption{Training Pipeline. The snow flake symbol denotes frozen parameters, the flame that a model is trained in a particular stage.}
\label{fig:overview_train}
\vspace{-15pt}
\end{figure}

\vspace{-4pt}
\subsection{Pre-Training on AudioSet strong}
\vspace{-4pt}

\label{subsec:as_pre-training}
We hypothesize that the transformer models would benefit from additional pre-training on a large dataset strongly annotated for various acoustic events. To this end, we add a BiGRU block with 1024 units that processes the output of the transformer. We pre-train for 10 epochs on AudioSet strong~\cite{hersheyasstorng}, a subset of AudioSet that holds around 86,000 strongly labeled examples with annotations for 456 event classes. While ATST and, in particular, fPaSST benefit from this pre-training, there was no effect on the downstream task performance of BEATs; therefore, we only pre-train ATST and fPaSST on AudioSet strong. We select the checkpoint with the highest PSDS1 score on the AudioSet strong validation set for downstream training.


\vspace{-4pt}
\subsection{Multi-Stage Training}
\vspace{-4pt}

Inspired by~\cite{shao2024fine}, \textit{I1} and \textit{I2} are both split into two training stages. In \textit{S1}, the CRNN (CNN + BiGRU) is trained from scratch while the large transformer model is frozen. This setup corresponds to the training of the baseline system with slightly different hyperparameters and additional data augmentations (as shown in Table~\ref{tab:aug}). 

In \textit{S2}, the CRNN is initialized with pre-trained weights from \textit{S1}, and both the CRNN and the transformer model are fine-tuned. As the system already performs well in its initial state, the transformer can rely on a high-quality self-supervised loss computed on the larger unlabeled set. Aligned with~\cite{shao2024fine}, in \textit{S2}, we compute the interpolation consistency (ICT) loss~\cite{zheng2021improved} in addition to the mean teacher (MT) loss~\cite{Tarvainen17mean_teacher}. In both stages, we choose the best model based on the validation set by computing the sum of PSDS1 on the strongly labeled synthetic data, PSDS1 on external strongly labeled real data, and mpAUC on the MEASTRO validation set.

\vspace{-4pt}
\subsection{Multi-Iteration Training}
\vspace{-4pt}

After completing \textit{I1}, we build an ensemble (see \textit{Ensemble Stage 2} in Table~\ref{tab:results}) of multiple ATST, fPaSST, and BEATs models. This ensemble is used to compute strong pseudo-labels for all audio clips in the training set by averaging the frame-wise logits of the individual models. In \textit{S1} of \textit{I2}, we then use the pseudo-labels as additional prediction targets. We found that BCE is superior to MSE for computing the pseudo-label loss, and interestingly, using the pseudo-label loss only improves performance in \textit{S1} of \textit{I2} (see Table~\ref{tab:abl}). 

\vspace{-4pt}
\section{Experimental Setup}
\label{sec:exp_setup}
\vspace{-4pt}

\subsection{Audio Pre-processing and Augmentation}
\vspace{-4pt}

For all models, we convert audio clips to 10 seconds in length at a 16 kHz sampling rate. For the CNN, we match the baseline settings and compute Mel spectrograms with 128 Mel bins using a window length of 128 ms and hop size of 16 ms. For the transformers, we use their original feature extraction pipelines~\cite{li2022atstf, koutini22passt, chen2022beats}.

Table~\ref{tab:aug} details all the data augmentation methods used in our training pipeline. In contrast to the baseline, we apply Cross-Dataset Mixup and Cross-Dataset Freq-MixStyle. That is, we mix audio clips from MAESTRO and DESED instead of keeping them separate. In the case of Mixup, we allow the loss to be calculated for all partially active classes, irrespective of the audio clip's dataset origin (see Section~\ref{subsec:handling_heterogeneous}). For Wavmix and Mixup, we mix the pseudo-labels accordingly.

\vspace{-4pt}
\subsection{Datasets and Optimization}
\vspace{-4pt}

We use the DESED~\cite{Turpault2019DCASE} and MAESTRO~\cite{Martinmorato2023maestro} datasets as provided for Task 4 in the DCASE 2024 challenge and, additionally, approximately 7,000 strongly annotated clips extracted from AudioSet strong according to~\cite{xiao2023sound}. We refer the reader to \cite{Schmid2024tech_repo} for a detailed description of the data setup.

The training data can be seen as the union of five subsets: MAESTRO strong and DESED: real strong, synthetic strong, weakly annotated, and unlabeled. We draw batches of (12, 10, 10, 20, 20) and (56, 40, 40, 72, 72)  samples from these datasets in \textit{S1} and \textit{S2}, respectively. The model is trained to minimize BCE loss on all (pseudo-)labeled audio clips and MSE loss for the self-supervised MT~\cite{Tarvainen17mean_teacher} and ICT~\cite{zheng2021improved}methods. We compute a weighted sum of all losses and tune the individual weights for all iterations and stages. AdamW~\cite{loshchilov2017decoupled} with weight decays of 1e-2 and 1e-3 is used in \textit{S1} and \textit{S2}, respectively. Learning rates are listed in Table~\ref{tab:results}.

\begin{table}[t]
\begin{center}
\begin{tabular}{@{}lccc@{}}
\toprule
\textit{Aug. Method} & \textbf{Target} & \textbf{HP} & \textbf{Pipeline} \\ \midrule
DIR~\cite{morocutti2023device} & All & $p$=0.5 & \textit{I\{1,2\}.S2} \\ 
Wavmix~\cite{zhang2017mixup} & Str. & $p$=0.5,$\alpha$=0.2 & \textit{I\{1,2\}.S\{1,2\}}  \\ \midrule
Freq-MixStyle~\cite{Kim22freqm} & All & $p$=0.5,$\alpha$=0.3 & \textit{I1.S\{1,2\}},\textit{I2.S2}\\ 
Mixup~\cite{zhang2017mixup}  & All & $p$=0.5,$\alpha$=0.2 & \textit{I\{1,2\}.S\{1,2\}} \\
Time-Masking & DES. Str. & $s$=[0.05,0.3] & \textit{I\{1,2\}.S2} \\ 
FilterAugment~\cite{nam2022filteraugment} & All & linear,p=0.8 & \textit{I1.S\{1,2\}},\textit{I2.S2}    \\ 
Freq-Warping~\cite{li2022atst} & All & p=0.5 & \textit{I\{1,2\}.S2}   \\  \bottomrule
\end{tabular}
\caption{The table lists data augmentation methods, the data subset they are applied to (\textbf{Target}), hyperparameters (\textbf{HP}), and the respective iteration and stage they are used in (\textbf{Pipeline}). \textit{p} is the probability for applying the augmentation method; $\alpha$ parameterizes Beta distributions; $s$ specifies the masking ratio interval; and (\textit{DES.}) \textit{Str.} refers to strongly annotated audio clips (from DESED).}
\label{tab:aug}
\end{center}
\vspace{-20pt}
\end{table}

\begin{table*}[h!]
\vspace{-14pt}
\centering
\begin{tabular}{@{}l|l|l|ccccc|ccc@{}}
\toprule
    &   & \textbf{Model}  & \textbf{lr\_cnn} & \textbf{lr\_rnn} & \textbf{lr\_tf} & \textbf{lr\_dec} & \textbf{Seq.} & \textbf{mpAUC} & \textbf{PSDS1}  & \textbf{Rank Score}\\ 
     
      \midrule
\multirow{ 6}{*}{Iteration 1} & \multirow{ 3}{*}{Stage 1} & ATST  & 1e-3 & 1e-3 & - & - & int. lin. & 0.702 $\pm$ 0.008 & 0.493 $\pm$ 0.012 & 1.195 $\pm$ 0.012 \\
 & & fPaSST  & 1e-3 & 1e-3 & - & - & int. nearest & 0.709 $\pm$ 0.021 & 0.502 $\pm$ 0.010 & 1.212 $\pm$ 0.027 \\
 &  & BEATs & 1e-3 & 1e-3 & - & - & int. nearest & 0.719 $\pm$ 0.004 & 0.509 $\pm$ 0.003 & \textbf{1.228 $\pm$ 0.006} \\ \cline{2-11}
& \multirow{ 3}{*}{Stage 2} & ATST  & 1e-4 & 1e-3 & 1e-4 & 0.5  & int. nearest & 0.739 $\pm$ 0.017 & 0.520 $\pm$ 0.005 & \textbf{1.259} $\pm$ 0.020 \\
&  & fPaSST & 1e-4 & 1e-3 & 1e-4 & 1 & int. nearest & 0.726 $\pm$ 0.021 & 0.514 $\pm$ 0.008 & 1.24 $\pm$ 0.027 \\
&  & BEATs & 1e-4 & 1e-3 & 1e-4 & 1 & int. lin. & 0.713 $\pm$ 0.002 & 0.539 $\pm$ 0.004 & 1.252 $\pm$ 0.003 \\ \cline{2-11}
& \multicolumn{2}{c}{Ensemble Stage 2} & - & - & - & - & mix & 0.735 & 0.569 &  1.303 \\\hline
\multirow{ 6}{*}{Iteration 2} & \multirow{ 3}{*}{Stage 1} & ATST  & 5e-4 & 5e-4 & -  & - & avg. pool & 0.741 $\pm$ 0.017 & 0.536 $\pm$ 0.006 & \textbf{1.277 $\pm$ 0.012} \\
&  & fPaSST & 5e-4 & 5e-4 & - & - & int. nearest & 0.722 $\pm$ 0.011 & 0.526 $\pm$ 0.004 & 1.248 $\pm$ 0.012 \\
&   & BEATs & 5e-4 & 5e-4 & - & - & int. nearest & 0.724 $\pm$ 0.011 & 0.537 $\pm$ 0.005 & 1.262 $\pm$ 0.010 \\ \cline{2-11}
 & \multirow{ 3}{*}{Stage 2} & ATST  & 1e-5 & 1e-4 & 1e-4 & 0.5 & avg. pool & 0.750 $\pm$ 0.004 & 0.548 $\pm$ 0.004 & \textbf{1.298} $\pm$ 0.006  \\
&  & fPaSST & 5e-5 & 5e-4 & 1e-4 & 1 & int. nearest & 0.719 $\pm$ 0.013 & 0.539 $\pm$ 0.003 & 1.259 $\pm$ 0.015  \\
&  & BEATs & 5e-5 & 5e-4 & 1e-4 & 1 & int. nearest & 0.729 $\pm$ 0.005 & 0.557 $\pm$ 0.005 & 1.286 $\pm$ 0.009 \\ 
\bottomrule
\end{tabular}
\caption{
 The table presents the results of ATST, fPaSST, and BEATs for both iterations and stages on the official development test set. For each model, we list the best configuration in terms of the sequence length adaptation method (\textbf{Seq.}), where \textit{int. lin.}, \textit{int. nearest}, \textit{avg. pool}, and \textit{mix} denote linear and nearest-exact interpolation, adaptive average pooling, and a mixture of these methods, respectively. \textit{Ensemble Stage 2} is used to generate the pseudo-labels for Iteration 2. \textbf{Rank Score} denotes the sum of \textbf{mpAUC} and \textbf{PSDS1}.
}
 \label{tab:results}
\vspace{-4pt}
\end{table*}

\begin{table}[t]
\begin{center}
\begin{tabular}{@{}c|cccc@{}}
\toprule
\textbf{Model} & \textbf{mpAUC} & \textbf{PSDS1 MF} & \textbf{PSDS1*} & \textbf{Ev. PSDS1*}  \\ \midrule
ATST & 0.750 & 0.548 & 0.617 & 0.684 \\
fPaSST & 0.719 & 0.539 & 0.601 & 0.681 \\ 
BEATs & 0.729 & 0.557 & 0.622 & 0.683 \\ \midrule
ATST DT & \xmark & \xmark & \xmark & 0.692 \\
\bottomrule
\end{tabular}
\caption{Results for best single-model configurations of ATST, fPaSST, and BEATs from \textit{I2.S2}. \textbf{PSDS1} lists results with a median filter; \textbf{PSDS1*} results using cSEBBs postprocessing~\cite{ebbers2024sound}; and \textbf{Ev. PSDS1*} lists results on the DESED public evaluation set with cSEBBs postprocessing. \textit{ATST DT} denotes the best ATST configuration trained on the full development set.}
\label{tab:sub}
\end{center}
\vspace{-20pt}
\end{table}

\vspace{-4pt}
\subsection{Handling Heterogeneous Sound Event Classes}
\label{subsec:handling_heterogeneous}
\vspace{-4pt}

The DESED and MAESTRO datasets are annotated with two different sets of sound event classes. We adopt the baseline~\cite{cornell2024dcase}  strategy, in which the loss for an audio clip is calculated only on the dataset-specific event classes and mapped event classes, as explained in the following: To exploit the fact that the DESED and MAESTRO classes are not fully disjoint but partly represent the same concepts, the baseline introduces class mappings. For example, when the classes \textit{people talking}, \textit{children voices}, or \textit{announcement} are active in a MAESTRO clip, the corresponding DESED class \textit{Speech} is set to the same confidence value. In addition, we also include a mapping from MAESTRO to DESED classes. Specifically, we set the values of the MAESTRO classes \textit{cutlery and dishes} and \textit{people talking} to 1 if the DESED classes \textit{Dishes} and \textit{Speech} are present. This is also performed for weak class labels.

\vspace{-4pt}
\subsection{Postprocessing}
\label{subsec:post}
\vspace{-4pt}

For model selection and hyperparameter tuning, we stick with the same class-wise median filter used in the baseline system~\cite{cornell2024dcase}. After selecting the best configurations for each model, we apply the recently introduced Sound Event Bounding Boxes (SEBBs)~\cite{ebbers2024sound} method for postprocessing. We use class-wise parameters and tune them by using linearly spaced search grids (8 values) for step filter length (0.38 to 0.66), relative merge threshold (1.5 to 3.25), and absolute merge threshold (0.15 to 0.325). 

\begin{table}[t]
\begin{center}
\begin{tabular}{@{}c|ccc@{}}
\toprule
\textbf{System} & \textbf{mpAUC} & \textbf{PSDS1} & \textbf{Rank Score}  \\ \midrule
\textbf{ATST \textit{I2.S1}}  & 0.741 & 0.536 & \textbf{1.277}  \\ \midrule
- DESED & 0.724 & -  & - \\
- MAESTRO & - & 0.531 & - \\ \midrule
- SSL MAESTRO & 0.741 & 0.535 & 1.276 \\
- MAESTRO-DESED Map. & 0.717 & 0.530 & 1.247  \\
+ SSL class mask & 0.740 & 0.530 & 1.27 \\
+ Separate RNN layer & 0.714 & 0.531 & 1.244 \\
+ Hard Pseudo & 0.706 & 0.538 & 1.244  \\
+ Pseudo All Classes & 0.717 & 0.534 & 1.25  \\ \midrule \midrule
\textbf{ATST \textit{I2.S2}}  & 0.750 & 0.548 & \textbf{1.298}  \\ \midrule
- SSL MAESTRO & 0.743 & 0.546 & 1.289  \\
- MAESTRO-DESED Map. & 0.749 & 0.547 & 1.297  \\
- SSL class mask & 0.749 & 0.544 & 1.293 \\
+ Pseudo Loss & 0.746 & 0.552 & 1.297  \\
\bottomrule
\end{tabular}
\caption{Ablation Study on design choices related to the heterogeneous datasets and the pseudo-label loss used in \textit{I2.S1}. The study is performed on the top single model, ATST, trained in \textit{I2.S1} (upper part) and in \textit{I2.S2} (lower part).}
\label{tab:abl}
\end{center}
\vspace{-20pt}
\end{table}

\vspace{-4pt}
\section{Results}
\label{sec:results}
\vspace{-4pt}

In this section, we present the results of the described models (Section~\ref{sec:arch}) in the introduced training pipeline (Section~\ref{sec:training_setup}). Table~\ref{tab:results} lists the best configuration and the corresponding results on the test set for each architecture in both iterations and stages. The table lists the sequence pooling method (\textbf{Seq.}) and the CNN (\textbf{lr\_cnn}), RNN (\textbf{lr\_rnn}), and Transformer (\textbf{lr\_tf}) learning rates. \textbf{lr\_dec} indicates the layer-wise learning rate decay for the transformers as used in~\cite{shao2024fine}. 

In \textit{I1.S1}, in which the transformers are frozen, BEATs seems to extract the embeddings of the highest quality, followed by fPaSST and ATST. \textit{I1.S1} with BEATs is very similar to the baseline~\cite{cornell2024dcase} and achieves a similar rank score with a slight performance increase in our setup. Compared to \textit{I1.S1}, all three transformers demonstrate a large increase in rank score when fine-tuned on the Task 4 datasets in \textit{I1.S2}. Notably, the three transformers have different strengths, with ATST and BEATs achieving the best scores on MAESTRO and DESED clips, respectively. \textit{Ensemble Stage 2} denotes an ensemble of 46 models resulting from \textit{I1.S2}, including ATST, fPaSST, and BEATs trained in different configurations. We use \textit{Ensemble Stage 2} to generate strong pseudo-labels for all audio clips in the dataset. 

The additional pseudo-label loss in \textit{I2.S1} boosts performance substantially, with all three transformers achieving a higher rank score compared to \textit{I1.S2}. The top rank scores for all models are achieved in \textit{I2.S2}, with ATST obtaining the highest rank score. 

Table~\ref{tab:sub} presents the top configurations of ATST, fPaSST, and BEATs from \textit{I2.S2} with the state-of-the-art postprocessing method cSEBBs~\cite{ebbers2024sound} applied. \textit{ATST} and \textit{ATST DT}, a variant of ATST that is trained on all available audio clips included in the Task 4 development set, were submitted as single models to the challenge. \textit{ATST DT} using cSEBBs postprocessing achieves a PSDS1 of 0.692 on the public evaluation set of DESED, improving over the previous state of the art (0.686 PSDS1)~\cite{ebbers2024sound}.




 \vspace{-4pt}
\subsection{Ablation Study}
\label{subsec:abl}
\vspace{-4pt}

Table~\ref{tab:abl} shows the results of ATST for \textit{I2.S1} and \textit{I2.S2} trained in different configurations to analyze the design choices related to the heterogeneous datasets and the pseudo-label loss. For settings \mbox{\textit{- DESED}} and \mbox{\textit{- MAESTRO}}, the proposed system is trained only on MAESTRO and DESED data, respectively. We find that training on DESED and MAESTRO simultaneously is beneficial for the performance on both datasets, which coincides with the finding reported for the baseline system~\cite{cornell2024dcase}. For both stages of \textit{I2}, excluding MAESTRO clips when calculating the self-supervised losses (\mbox{\textit{- SSL MAESTRO}}) and not mapping event classes from MAESTRO to DESED (\mbox{\textit{- MAESTRO-DESED Map.}}, see Section~\ref{subsec:handling_heterogeneous}) leads to a performance decrease. However, we find no clear answer to the question of whether the SSL loss should be calculated on all classes or only on the dataset-specific classes of an audio clip (\textit{+/- SSL class mask}); \textit{S1} and \textit{S2} benefit from different settings. Interestingly, using the pseudo-label loss in \textit{I2.S2} (\textit{+~Pseudo Loss}) does not increase the rank score. Therefore, the setup in \textit{I1.S2} and \textit{I2.S2} remains identical, which demonstrates that a well-trained CRNN from \textit{S1} can have a large impact on the performance achieved in \textit{S2}. We also tried to use separate heads for predictions on DESED and MAESTRO classes and realized this with an additional single BiGRU layer per dataset (\textit{+~Separate RNN layer}), which resulted in a performance decrease. Further obvious choices, such as thresholding the pseudo-labels by 0.5 (\textit{+~Hard Pseudo}) and calculating the pseudo-label loss on all classes (\textit{+~Pseudo All Classes}) instead of only dataset-specific classes, are inferior to our proposed strategy as well.

\vspace{-4pt}
\section{Conclusion}
\label{sec:conclusion}
\vspace{-4pt}

This paper presented a multi-iteration, multi-stage training routine for fine-tuning transformers on the SED task with heterogeneous datasets. We showed that the performance of all tested systems monotonously increases throughout both iterations and stages. The proposed method led to a new state-of-the-art performance of 0.692 in PSDS1 on the DESED public evaluation set and achieved the top rank in Task 4 of the DCASE 2024 challenge. We specifically studied design choices related to the heterogeneous datasets and found that simultaneously training on DESED and MAESTRO leads to a performance increase on both datasets compared to training the system on a single dataset.

\vspace{-4pt}
\section{ACKNOWLEDGMENT}
\label{sec:ack}
\vspace{-4pt}

The computational results presented were achieved in part using the Vienna Scientific Cluster (VSC) and the Linz Institute of Technology (LIT) AI Lab Cluster. The LIT AI Lab is supported by the Federal State of Upper Austria. Gerhard Widmer's work is supported by the European Research Council (ERC) under the European Union's Horizon 2020 research and innovation programme, grant agreement No 101019375 (Whither Music?).

\bibliographystyle{IEEEtran}
\bibliography{refs}

\end{sloppy}
\end{document}